\documentclass[twocolumn,aps,prl,raggedbottom,nobalancelastpage,amssymb,superscriptaddress,showpacs]{revtex4-1}

\usepackage{amsmath}
\usepackage{amssymb}
\usepackage{amsfonts}
\usepackage{dsfont}
\usepackage{graphicx}
\usepackage{bm}
\usepackage{appendix}
\usepackage{epsfig}

\newcommand{\ket}[1]{\left|{#1}\right>}

\usepackage[usenames,dvipsnames]{color}
\usepackage{ulem}
\usepackage{soul}
\setstcolor{red}
\sethlcolor{yellow}

\begin{document}
\title{A scattering matrix formulation of the topological index of
  interacting fermions in one-dimensional superconductors}

\author{Dganit Meidan}
\affiliation{\mbox{Department of Physics, Ben-Gurion University of the Negev, BeÕer-Sheva 84105, Israel}}
\affiliation{\mbox{Dahlem Center for Complex Quantum Systems and Fachbereich Physik, Freie Universit\"at Berlin, 14195 Berlin, Germany}}
\author{Alessandro Romito}
\affiliation{\mbox{Dahlem Center for Complex Quantum Systems and Fachbereich Physik, Freie Universit\"at Berlin, 14195 Berlin, Germany}}
\author{Piet W. Brouwer}
\affiliation{\mbox{Dahlem Center for Complex Quantum Systems and Fachbereich Physik, Freie Universit\"at Berlin, 14195 Berlin, Germany}}

\begin{abstract}
We construct a scattering matrix formulation for the topological classification of one-dimensional
 superconductors with effective time reversal symmetry in the presence of interactions. For a closed geometry,
Fidkowski and Kitaev have shown that such systems have a $\mathbb{Z}_8$ topological classification. 
We show that in the weak coupling limit, these systems retain a unitary scattering matrix at zero 
temperature, with a topological index given by the trace of the Andreev reflection matrix, $\mbox{tr}\, r_{\rm he}$. With interactions, $\mbox{tr}\, r_{\rm he}$ generically takes on the finite set
of values $0$, $\pm 1$, $\pm 2$, $\pm 3$, and $\pm 4$. We show that  the two topologically equivalent 
phases with $\mbox{tr}\, r_{\rm he} = \pm 4$  support  emergent {\it many-body} end states,
which we identify to be a topologically protected Kondo-like resonance. The path in phase space that 
connects these equivalent phases crosses a non-fermi liquid fixed point where a multiple channel 
Kondo effect develops. 
Our results  connect the topological index to transport properties, thereby highlighting the experimental 
signatures of interacting topological phases in one dimension.
\end{abstract}
\pacs{
}

\maketitle

\paragraph{Introduction} 

Superconducting wires exist in two topologically distinct classes, which can be distinguished 
by the presence or absence of a Majorana bound state at the wire's ends \cite{Hasan2010,Qi2011,Alicea2012,Stanescu2013}. 
Because of their non-Abelian exchange 
statistics \cite{Moore91,Read2000,Ivanov2001} such Majorana states have been proposed as an element 
for a fault-tolerant quantum computational scheme \cite{Kitaev2003,Kitaev2006,Freedman1998,Nayak2008}. 
The topological phases have been shown to be robust against the presence of (moderate)
disorder \cite{Motrunich2001,Brouwer2011} or interactions \cite{SGangadharaiahPRL2011,LFidkowskiPRB2011c,EMStoudenmirePRL2011,ESelaPRB2011}. While only few rather exotic
systems are believed to realize the nontrivial phase \cite{Moore91,GEVolovik2003,SDSarmaPRB2006}, 
such systems can in principle be engineered in solid state devices \cite{Fu2008,Lutchyn2010,Oreg2010}. 
Recent experiments on semiconductor wires proximity coupled to superconductors have been reported
to show indications of the existence of Majorana bound states \cite{Mourik2012,Das2012}.

Whereas the topological classification of superconducting wires with or without Majorana states
exclusively rests on the presence of particle-hole symmetry, in many of the proposals for the
actual realization of topological superconducting wires an additional approximate {\it effective time-reversal} 
symmetry appears \cite{Tewari2011}. For the $p+ip$ model of a spinless superconductor wire of
width $W$ and Fermi velocity $v_{\rm F}$, the effective time reversal
symmetry requires the superconducting gap  to be much smaller than the
transverse quantization energy $\hbar v_{\rm F}/W$
\cite{GKellsPRB2012a},  
a condition that is easily met in view of the generic smallness of the
gap in the proposals
to engineer topological superconductors.
With effective time reversal symmetry, Majorana states carry a sign, such that Majorana
states of the same sign coexist at the same end of the superconducting wire. As a result, the effective time reversal symmetry
symmetry changes the topological classification from $\mathbb{Z}_2$ to $\mathbb{Z}$, which counts the number of Majorana end states, with sign.

Unlike the $\mathbb{Z}_2$ classes, the $\mathbb{Z}$ effective time reversal symmetry classification is not stable against the
presence of interactions, even if the interactions preserve the symmetry. In a seminal work,
Fidkowski and Kitaev \cite{Fidkowski2010} showed that interactions break the free-fermion $\mathbb{Z}$ 
topological index down to $\mathbb{Z}_8$ (see also Refs.
\onlinecite{Turner2011,Fidkowski2011a,Gurarie2011,Manmana2012}). To understand this result, one notes
that for the seven topological classes with $n=0,\pm 1, \pm 2$, and
$\pm 3$ there exist realizations with less than two fermions at each
wire's end (as two Majoranas combine into one fermion). Hence, no
local interaction term is allowed for the low-energy sector and the free-fermion classification remains 
valid. Fidkowski and Kitaev then showed that the two classes with $n=\pm 4$ can be adiabatically 
connected by a suitable interaction, so that eight inequivalent classes remain \cite{note1}. 

The experimental observation of Majorana states inevitably relies on the coupling of the topological 
superconductor to a normal-metal contact, the simplest detection scheme being through a measurement of
the two-terminal Andreev conductance \cite{Mourik2012,Das2012}. 
With effective time reversal  symmetry, but without interactions,  the 
Andreev conductance $G$ at zero bias and zero temperature takes the quantized value 
$G= 2 |n| e^2/h$, if there are $n$ Majorana end states~\cite{Law2009,Flensberg2010,note2}. Distinguishing 
positive and negative $n$ requires a phase-sensitive measurement of the Andreev reflection matrix, using $\mbox{tr}\, r_{\rm he} = n$~\cite{Fulga2011}.

What is the Andreev-reflection signature of the $\mathbb{Z}_8$ topological classification for a
superconducting wire in the effective time reversal symmetry class,
and what is the nature of the emergent end states if the wire is
weakly coupled to a normal contact? In this letter we answer these questions, showing how the 
breakdown of the free-fermion $\mathbb{Z}$ classification to $\mathbb{Z}_8$ is reflected in the 
Andreev reflection matrix $r_{\rm he}$. 
Establishing the scattering properties of interacting topological phases highlights the experimental 
signatures of the emergent end states in conductance measurements or Josephson currents, thereby 
providing an important link between recent theoretical advances and future experimental work.  

Our main results can be summarized as follows: For a normal lead weakly coupled to the 
superconducting wire, we find (i) that generically the scattering matrix is well defined and unitary
at zero temperature,
in spite of the presence of interactions in the superconducting wire, and (ii) that in the presence of interactions, $\mbox{tr}\, r_{\rm he}$
is restricted to the values $0$, $\pm 1$, $\pm 2$, $\pm 3$, and $\pm 4$. Exceptions to these rules do occur, but they form a set of measure zero and are unstable 
to perturbations. The topological phases with $-3 \le \mbox{tr}\, r_{\rm he} \le 3$ are topologically
equivalent to their free-fermion analogues with $|n| \le 3$ uncoupled Majorana end states. 
Conversely, we show that the phases with $\mbox{tr}\, r_{\rm he} = \pm 4$ are characterized by 
emergent {\it many-body}  end states, which we identify as topologically protected Kondo-like 
resonances. These two configurations, which are both stable with respect to perturbations,
are topologically {\it equivalent} in the presence of interactions. The path in phase space that 
connects these equivalent phases crosses a non-fermi liquid fixed point associated with a multiple 
channel Kondo problem, without closing the bulk excitation gap in the superconductor.

\paragraph{Model.}
We derive these results in the framework of a multi-channel Majorana chain~\cite{Kitaev2001} with 
an effective time reversal symmetry, $\mathcal{T}^2=1$, corresponding to class BDI in the Cartan 
classification~\cite{Altland1997}. 
In the absence of interactions, the Majorana chain is described by the  Hamiltonian
\begin{align}
  H_{\rm S} = &   \sum_{\alpha,j} + (\Delta_{\alpha} d_{j+1,\alpha}^\dag
  d_{j,\alpha}^\dag - t d_{j+1,\alpha}^\dag d_{j,\alpha }  \!+\!{\rm
    h.c. }) &  \nonumber \\
 & + \sum_{\alpha,j} \mu \, d_{j,\alpha}^\dag d_{j,\alpha}, &
  \label{eq:H0}
\end{align}
where $t>0$ is the hopping parameter, $\mu$ the chemical potential, and $d_{j,\alpha}$ is the annihilation 
operator for a fermion on site $j$ and channel $\alpha$. We consider
$j=1,2,\ldots$, so that Eq.\ (\ref{eq:H0}) describes a half-infinite wire beginning at $j=1$.
We choose $\Delta_{\alpha} \neq 0$ to be real, so that $H_{\rm S}$ is invariant under an effective
antiunitary time-reversal symmetry operation ${\cal T}$, $\mathcal{T} d_{j,\alpha} \mathcal{T}^{-1} = 
d_{j,\alpha}$.

Equation (\ref{eq:H0}) can be conveniently rewritten in terms of Majorana operators,
$\gamma_{j,\alpha} =d_{j,\alpha}+d_{j,\alpha}^\dag$ and
$\tilde{\gamma}_{j,\alpha} =-i(d_{j,\alpha}-d_{j,\alpha}^\dag)$. 
The model undergoes a topological phase transition at $|\mu|=2 |t|$. The topological phase is 
characterized by two zero energy Majorana end states for each channel, exponentially localized at 
opposite ends of the chain and separated by an energy gap from the bulk excitations. Since we consider
a half-infinite chain, only the end state located near $j=1$ is of relevance for us. For positive 
$\Delta_{\alpha}$ the operator for this Majorana end state commutes with ${\cal T}$ and we denote
it by $\gamma_{\alpha}$; for negative $\Delta_{\alpha}$ the end-state operator anticommutes with 
${\cal T}$, and we write $\tilde{\gamma}_{\alpha}$.
The effective time-reversal symmetry permits perturbations that couple Majorana end states of different type,
$\mathcal{T} i\gamma_{\alpha} \tilde{\gamma}_\beta \mathcal{T}^{-1} = i \gamma_\alpha \tilde{\gamma}_\beta$, 
but forbids coupling between Majorana end modes of the same type. The system is therefore classified by a 
$\mathbb{Z}$ topological index, which counts the number $N$ of Majorana end states of ``$\gamma$'' type minus 
the number $\tilde N$ of end states of ``$\tilde \gamma$'' type. Without loss of generality, we take the
channels $\alpha=1,2,\ldots,N$ to be of ``$\gamma$'' type, whereas the remaining channels 
$\alpha=N+1,\ldots,N + \tilde N$ are of ``$\tilde \gamma$'' type.

\paragraph{Scattering matrix classification.} 
To study the scattering properties, we couple the semi-infinite wire to a half-infinite normal 
lead. The Hamiltonian is given by
\begin{align}
  H = H_{\rm S} + H_{\rm lead} + H_{\rm T},
\label{eq:Hnonint}
\end{align}
where $H_{\rm lead}$ is the Hamiltonian of the non-interacting ideal lead, $
  H_{\rm lead} = \sum_{k,\alpha} \xi_{k} c_{k,\alpha}^\dag c_{k,\alpha}$.
Here $c_{k,\alpha}$ is the annihilation operator for an electron with momentum $k$ 
measured with respect to the Fermi point and channel index $\alpha$ and $\xi_{k}$ the corresponding
kinetic energy. The term $H_{\rm T}$ describes tunneling between the lead and 
the superconductor, $ H_{\rm T}= \tilde{t}_{\rm T} \sum_{k,\alpha}
  c_{k,\alpha}^\dag d_{1,\alpha}+ \mbox{h.c.}$,
$\tilde{t}_{\rm T}$ being the tunneling matrix element. Considering that the superconducting wire is gapped in the bulk, we can project the tunneling Hamiltonian 
onto the low-energy sector consisting of the zero-energy Majorana end-states,
\begin{equation}
  H_{\rm T} = \frac{t_{\rm T}}{2} \sum_{k} \left(
  \sum_{\alpha=1}^{N} c_{k,\alpha}^\dag \gamma_{\alpha} + 
  i \sum_{\alpha=N+1}^{N + \tilde{N}} c_{k,\tilde{\alpha}}^\dag 
  \tilde{\gamma}_{\tilde{\alpha}} \right) ,
\end{equation}
where $t_T$ is the effective tunneling coupling of the end-state
Majorana to the lead.
As the system is gapped, no quasiparticle excitations are transmitted through the superconductor, and 
scattering processes are described by a unitary reflection matrix. 
In the BDI class, each Majorana end mode gives rise to perfect Andreev reflection, although the sign
of the reflection amplitude is opposite for the two types of Majorana modes. The topological index 
is  $n \equiv N - \tilde N = \mbox{tr}\, r_{\rm he}$~\cite{Fulga2011,Rieder2013} ,
where  $r_{\rm he}$ is evaluated at zero energy.

\paragraph{Generalization to interacting systems.}
We now generalize the above classification scheme to interacting systems. We consider short-range
two-fermion interactions, so that, after projection to the low-energy sector of the Majorana end
states, the interaction Hamiltonian involves the end-state operators $\gamma_{\alpha}$ and 
$\tilde \gamma_{\alpha}$ only~\cite{Note1,Beri2012,Altland2013}. If all Majorana end states are of the ``$\gamma$'' type, the most 
general interaction  in the low-energy sector has the form
\begin{eqnarray}
  H_{\rm int}=\sum_{\alpha_1 <\alpha_2<\alpha_3<\alpha_4}
  W_{\alpha_1 \alpha_2\alpha_3 \alpha_4} \gamma_{\alpha_1}
\gamma_{\alpha_2}\gamma_{\alpha_3} \gamma_{\alpha_4}.
  \label{eq:HW}
\end{eqnarray}
This interaction Hamiltonian commutes with the effective time-reversal operator ${\cal T}$.
If there are end states of both types, additional terms with four operators $\tilde \gamma$ or mixed
terms with two operators of each type are also allowed by the effective time-reversal symmetry.

We now discuss how the inclusion of the interaction Hamiltonian affects a Majorana chain with 
free-fermion topological index $n$. We'll find that the cases $|n| < 4$, $|n|=4$, 
and $|n| > 4$ are qualitatively different, and discuss these three cases separately.

\paragraph{The case $|n| < 4$.} Since generic potential perturbations gap out pairs of Majorana
states of opposite type, for weak interactions it is sufficient to limit our discussion to the 
``minimal'' realizations of the topological phases, which have $\tilde N=0$, if the topological 
index $n = N$ is positive, and $N = 0$, if $n = -\tilde N$ is negative. 
Since the low-energy interaction Hamiltonian $H_{\rm int} = 0$ if $N < 4$ 
and $\tilde N = 0$, or if $\tilde N < 4$ and $N=0$, see Eq.\ (\ref{eq:HW}),
we immediately conclude that the Andreev-reflection
signatures of these phases are unaffected by interactions, so that the reflection matrix is unitary and
$\mbox{tr}\, r_{\rm he} = n$.

\paragraph{The case $|n| = 4$.}
The case $n=\pm 4$ is nontrivial, since already in its minimal realization it allows for a 
nontrivial interaction Hamiltonian $H_{\rm int} = W \gamma_1 \gamma_2 \gamma_3 \gamma_4$.
Introducing the fermionic operators $f_\uparrow=(1/2) (\gamma_1 +i\gamma_2)$, $f_\downarrow=(1/2)
(\gamma_3+i \gamma_4)$, the low-energy interaction Hamiltonian for the case $n=4$
takes the form $H_{\rm int} = -W ( 2f_\uparrow^\dag f_\uparrow -1) (2 f_\downarrow^\dag f_\downarrow -1)$.
The interaction lifts the four-fold degeneracy of the zero energy level of the free fermion system 
and creates two doubly degenerate correlated states at energy $\pm W$. 
The situation with a degenerate level located at the end of the superconducting wire 
closely resembles a local impurity problem and much insight can be gained by performing a mapping to 
the latter. Hereto, the tunneling Hamiltonian is rewritten using the operators 
$g_{{\rm L},\uparrow,k} = 2^{-1/2}( c_{k,1} + i c_{k,2} )$,
$g_{{\rm R},\uparrow,-k} =2^{-1/2} ( -c_{k,1}^\dag - i c_{k,2}^\dag )$,
$g_{{\rm L},\downarrow,k} =2^{-1/2} ( c_{k,3}^\dag + i c_{k,4} )$,
$g_{{\rm R},\downarrow,-k} =2^{-1/2} ( - c_{k,3}^\dag - i c_{k,4}^\dag )$, 
so that the resulting Hamiltonian takes the form of the symmetric
Anderson model~\cite{foot3}
\begin{eqnarray}\nonumber
  H &=&  \sum_{\beta={\rm L},{\rm R}} 
         \sum_{k,\sigma=\uparrow,\downarrow} 
  \left[ \xi_k 
  g_{\beta,\sigma,k}^\dag g_{\beta,\sigma,k}
  +
  \frac{t_{\rm T}}{2}  
  \left( g_{\beta,\sigma, k}^\dag f_\sigma +{\rm h.c.}\right) 
  \right] \\
 && \mbox{} - W(2f_\uparrow^\dag f_\uparrow-1)(2f_\downarrow^\dag f_\downarrow-1).
  \label{eq:HamSymmAnd}
\end{eqnarray}
In general, the scattering matrix of this Kondo-like problem is non-unitary due to inelastic 
spin flip processes. However, at zero temperature unitarity is recovered due to the formation of a Kondo 
screening cloud, and the scattering matrix takes the simple form~\cite{Ng1988}
\begin{align}
  S_{\sigma,\sigma'} = \delta_{\sigma,\sigma'} \left.\left(
\begin{array}{ccc}
  r_{\rm RR}  & t_{\rm RL}    \\
  t_{\rm LR} & r_{\rm LL}   
\end{array}
\right)\right |_\sigma=\left(
\begin{array}{ccc}
0  & -1    \\
 -1 & 0    
\end{array}
\right).
\end{align}
Returning to the original basis with the lead operators $c_{k,\alpha}$, one finds that
the normal reflection matrix $r_{\rm ee} = 0$,
whereas  $r_{\rm he} = \openone_{4\times 4}$. We conclude that at zero 
temperature the topological index follows the non-interacting formula $\mbox{tr} r_{\rm eh} \!=
n = 4$. As a physical consequence, a tunneling conductance experiment will show a zero-bias
zero-temperature Andreev conductance $G = 8e^2/h$, which is the same conductance as without interactions.
However, unlike in the non-interacting case, this quantized conductance peak is a consequence of a 
formation of a many-body
Kondo-like resonance at the end of the interacting superconductor, and it no longer signals the presence
of four Majorana states. We can interpret this collective 
state, which is pinned to the Fermi energy, as the emergent edge state of the interacting superconductor.

The same analysis can be applied to the $n = - 4$ case. 
It gives $r_{\rm he} = -\openone_{4\times 4}$ and $\mbox{tr} r_{\rm he} = n = -4$. This implies
that $\mbox{tr} r_{\rm he}$ takes different quantized values for $n=4$ and $n=-4$.
How should this result be interpreted in light of the knowledge that,
with interaction, classes with
$n=4$ and with $n=-4$ are topologically equivalent  \cite{Fidkowski2010}? 

\begin{figure}
\includegraphics[width=0.45 \textwidth]{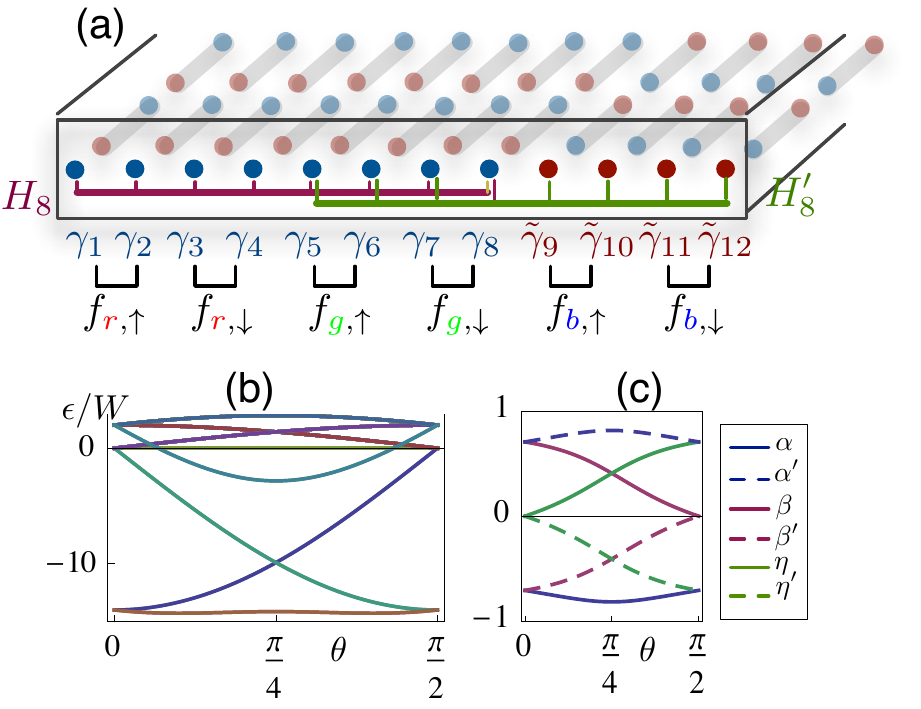}
\caption{ (color online) (a) Schematic picture of a twelve Majorana chains that interpolate between effective 
free-fermion classes $n=4$ and $n=-4$. The interactions terms coupling
different Majorana end-states are indicated by thick
lines. The``colored'' fermions constructed with Majorana fermions are
indicated explicitly.
(b) Energy spectrum as a function of the
interpolation parameter $\theta$, showing that the spectrum remains
gapped through the interpolation.
The ground state is two-fold degenerate. 
(c) Amplitudes $\alpha$, $\alpha'$, $\beta$, $\beta'$, $\gamma$, and $\gamma'$ for the ground state
wavefunctions, as a function of the interpolation paramter $\theta$, see Eq.\ (\ref{GSmanifold}).\label{fig2}
}
\end{figure} 

To find a scenario that resolves this paradox we monitor the Andreev reflection matrix along a 
path that connects the two classes $n=4$ and $n=-4$. The minimum 
channel number for a continuous interpolation between $n=4$ and $n=-4$ is twelve. 
For definiteness, we consider eight channels with positive $\Delta$,
labeled $\alpha=1,2,\dots,8$, and four channels with negative $\Delta$, labeled $\alpha=9,10,11,12$,
so that $N = 8$ and $\tilde N = 4$, see Fig.~\ref{fig2}.
For the interaction Hamiltonian we choose 
\begin{align}\label{crossoverH}
H_{\rm int}({\theta}) = W (H_8 \sin \theta  + H_8' \cos \theta )
\end{align} 
where
\begin{eqnarray}
\label{7to-1}
H_{8}&=&  \gamma_1 \gamma_2 \gamma_3 \gamma_4 
+ \gamma_5 \gamma_6 \gamma_7 \gamma_8 
+ \gamma_1 \gamma_2 \gamma_5 \gamma_6 
+ \gamma_3 \gamma_4 \gamma_7 \gamma_8 \\
\nonumber
&& \mbox{} - \gamma_2 \gamma_3 \gamma_6 \gamma_7 
- \gamma_1 \gamma_4 \gamma_5 \gamma_8
+ \gamma_1 \gamma_3 \gamma_5 \gamma_7
+ \gamma_3 \gamma_4 \gamma_5 \gamma_6\\
\nonumber
&& \mbox{} + \gamma_1 \gamma_2 \gamma_7 \gamma_8
- \gamma_2 \gamma_3 \gamma_5 \gamma_8 
 - \gamma_1 \gamma_4 \gamma_6 \gamma_7
+ \gamma_2 \gamma_4 \gamma_6 \gamma_8\\
\nonumber
&& \mbox{} - \gamma_1 \gamma_3 \gamma_6 \gamma_8
- \gamma_2 \gamma_4 \gamma_5 \gamma_7,
\end{eqnarray}
and $H_8'$ is obtained from $H_8$ by the substitution $\gamma_{\alpha} \to \gamma_{\alpha+4}$ for
$\alpha=1,2,3,4$ and $\gamma_{\alpha} \to \tilde \gamma_{\alpha+4}$ for $\alpha=5,6,7,8$. As shown by
Fidkowski and Kitaev, the Hamiltonian $H_8$ drives the eight Majorana end states in the first eight channels
into a nondegenerate topologically trivial ground state \cite{Fidkowski2010}. 
Similarly, the Hamiltonian $H_8'$ 
drives the eight Majorana states in the last eight channels into a nondegenerate trivial ground state.
(Since the last eight channels have four Majorana states of each type, a potential term coupling the 
last eight channels would have the same effect.) For $\theta=0$ the system is effectively in the 
free-fermion $n=4$ class analyzed previously, with  low-energy interaction Hamiltonian
$H_{\rm int} = W \gamma_1 \gamma_2 \gamma_3 \gamma_4$, while for $\theta=\pi/2$ it is in the opposite
case $n=-4$, with low-energy interaction Hamiltonian $W \tilde \gamma_{9} \tilde \gamma_{10} \tilde 
\gamma_{11} \tilde \gamma_{12}$. Thus, the family of interaction Hamiltonians 
$H_{\rm int}({\theta})$ with $0 \le \theta \le \pi/2$ smoothly interpolates between the (free-fermion) 
classes $n = 4$ and $n=-4$.

To continue our analysis, we construct six fermionic operators out of the twelve Majorana operators, 
which we group into three ``colors'' (r,g,b),
$f_{{\rm r},\uparrow} =  (1/2) (\gamma_{1} +i \gamma_{2})$, $f_{{\rm r},\downarrow}
= (1/2) (\gamma_{3} + i \gamma_{4})$, $f_{{\rm g},\uparrow} =  (1/2) (\gamma_{5} +i \gamma_{6})$, 
$f_{{\rm g},\downarrow} = (1/2) (\gamma_{7} + i \gamma_{8})$, $f_{{\rm b},\uparrow} =  (1/2) 
(\tilde \gamma_{9} +i \tilde \gamma_{10})$, $f_{{\rm b},\downarrow} = (1/2) (\tilde \gamma_{11} + 
i \tilde \gamma_{12})$, see Fig.\ \ref{fig2}. 
Diagonalization of the interaction Hamiltonian \eqref{crossoverH} reveals that the many-particle
ground state is two-fold degenerate for all $0 < \theta < \pi/2$ and confirms that the excitation
spectrum remains gapped otherwise, see Fig.\ \ref{fig2}. 
The two-fold degenerate ground state is spanned by the two states of the form
\begin{align}\label{GSmanifold}
\nonumber
\ket{\psi}&=\alpha \ket{111111} + \beta \ket{110000} + \eta \ket{000011}\\
\ket{\psi'} &=
\alpha' \ket{000000} + \beta'  \ket{001111} + \eta' \ket{111100},
\end{align}
where we use the basis of the occupations numbers $\ket{n_{r\uparrow}n_{r\downarrow}
n_{g\uparrow}n_{g\downarrow}n_{b\uparrow}n_{b\downarrow}}$. 
The excitation gap and the real coefficients $\alpha$, $\alpha'$, $\beta$, $\beta'$, $\gamma$, and
$\gamma'$ as a function of $\theta$ are shown in Fig.\ \ref{fig2}.
For $\theta \rightarrow 0^+$, we have $\eta=\eta' = 0$, and a transition between the two ground states 
is possible only by the exchange of two ``red''  
fermions with two lead fermions in the first four channels $\alpha=1,2,3,4$. In this case the system 
can be mapped to the symmetric Anderson model, as discussed above. Hence, at zero temperature, the 
reflection matrix is unitary and satisfies $\mbox{tr}\, r_{\rm eh} =4$.
Conversely, for $\theta \rightarrow \pi/2^-$, we find $\beta=\beta'=0$, and a transition between the two ground
states is possible only by the exchange of two ``blue'' fermions with lead channels in the last
set of four channels, $\alpha=9,10,11,12$. In this case
the system is again mapped to a symmetric Anderson model, but with $\mbox{tr}\, r_{\rm eh} =-4$ at 
zero temperature. For $0 < \theta < \pi/2$, generically all six amplitudes in Eq.\
(\ref{GSmanifold}) are nonzero. In this case transitions between the two ground states can take
place by the exchange of two fermions of arbitrary but {\it equal} colors. Since there are three colors
in total, the model is mapped to a three-channel Kondo problem. Such a multi-channel Kondo 
problem is, however, unstable, and the system flows to a single-channel Kondo fixed point determined 
by the strongest coupling constant, and correspondingly to one of the two extreme limits discussed 
above. Upon increasing $\theta$ from $0$ to $\pi/2$ invariably there must be a point at which
the coupling to the ``blue'' channels and the ``red'' or ``green'' channels is equal. This point is 
associated with 
a sharp phase boundary that exhibits a non-fermi liquid behavior, due to the formation of a multiple 
(generically: two) channel Kondo state at the wire's end. 
It follows that the scattering matrix goes through a non-unitary point along the path connecting these 
two phases. It is at this point that the transition between the quantized values $\mbox{tr}\, r_{\rm he} = 
\pm 4$ can take place. We note that this observation is consistent 
with recent studies that show that the crossover between topologically distinct non-interacting classes 
that become equivalent when interactions are present is associated with zeros of the Green's function 
indicating the formation of a non-fermi liquid state~\cite{Manmana2012}.


\paragraph{The case $|n| > 4$.}
Upon including interactions, the free-fermion cases with $|n| = 5$, $6$, $7$, or $8$ 
can be continuously connected to free-fermion classes with $|n| = 3$, $2$, $1$, and $0$,
respectively. An explicit example of a generic low energy Hamiltonian that  interpolates  between the free-fermion phases of $n=7 $ and $n=-1 $ is presented in the supplementary material. An analysis of the transitions between the two degenerate ground states reveals that the free-fermion $n=7$ configuration is unstable to interactions, and that in the 
weak-coupling limit $|t_{\rm T}| \ll |t|$  the interacting system  flows to an effectively non-interacting 
configuration with $\mbox{tr}\, r_{\rm he} = -1$. A similar analysis can be applied to $|n|=5$, $|n|=6$, and $|n|=8$.
\paragraph{Conclusions.}
We have constructed a scattering matrix formulation for the topological index of interacting fermions 
in one dimension with an effective time reversal symmetry. The scattering matrix of the
interacting system is unitary at zero temperature and zero energy, and the topological index can be
calculated from the trace $\mbox{tr}\, r_{\rm he}$. With interactions,
the topological index is restricted to nine possible values, $\mbox{tr}\, r_{\rm he} = 0$, $\pm 1$,
$\pm 2$, $\pm 3$, and $\pm 4$. Whereas the phases with $|\mbox{tr}\, r_{\rm he}| < 4$ are
effectively single-particle phases, we have shown that the two topologically equivalent phases with 
$\mbox{tr}\, r_{\rm he} = \pm 4$ are characterized by emergent {\it many-body} end states, which 
we identify to be a topologically protected Kondo-like resonance. The path in phase space 
that connects these equivalent phases crosses a non-fermi liquid fixed point where a multiple channel 
Kondo effect develops. 

Although the main motivation for our work is fundamental, a theory of implications of the 
topological classification for scattering properties is essential for a theoretical description 
of experimental geometries, where topological superconductors necessarily need to be connected to 
normal-metal probes. The effective time-reversal symmetry that is behind the $\mathbb{Z}_8$ topological
classification is relevant for some of the recent proposals to realize topological superconductors 
in one dimension. Whereas interactions are believed to be of minor importance if the topological 
superconductivity derives from the proximity of a bulk superconductor, experimental efforts to
minimizing the electric screening of nearby superconductors in order to allow for local 
gating of the device inevitably lead to a larger role of interactions in the topological
superconductor.

\paragraph{Acknowledgments} We gratefully  acknowledge discussion with Yigal Meir. This work is supported 
by the Alexander von Humboldt Foundation.


\begin{thebibliography}{99}
\expandafter\ifx\csname natexlab\endcsname\relax\def\natexlab#1{#1}\fi
\expandafter\ifx\csname bibnamefont\endcsname\relax
  \def\bibnamefont#1{#1}\fi
\expandafter\ifx\csname bibfnamefont\endcsname\relax
  \def\bibfnamefont#1{#1}\fi
\expandafter\ifx\csname citenamefont\endcsname\relax
  \def\citenamefont#1{#1}\fi
\providecommand{\bibinfo}[2]{#2}


\bibitem{Hasan2010}
M.~Z.~Hasan and C.~L.~Kane, 
\newblock Rev. Mod. Phys {\bf 82} 3045 (2010).

\bibitem{Qi2011}
X.-L.~Qi and S.-C.~Zhang,
\newblock Rev. Mod. Phys. {\bf 83} 1057 (2011).


\bibitem{Alicea2012}
J.~Alicea ,
\newblock Rep. Prog. Phys. {\bf 75}, 076501 (2012).
  
\bibitem{Stanescu2013}
T.~D.~Stanescu and  S.~Tewari,
\newblock J. Phys.: Condens. Matter {\bf 25}, 233201 (2013).

  \bibitem{Moore91} G.\ Moore, N.\ Read, Nucl.\ Phys.\ B {\bf 360}, 362 (1991).

\bibitem{Read2000}
N.~Read and D.~Green,
\newblock Phys. Rev. B {\bf 61}, 10267 (2000).

\bibitem{Ivanov2001}
D. Ivanov, 
\newblock Phys. Rev. Lett. {\bf86}, 268 (2001). 

\bibitem{Kitaev2003}
A.~Kitaev,
\newblock Ann. Phys. {\bf 303}, 2 (2003).

\bibitem{Freedman1998}
M.~H. Freedman,
\newblock Proc. Natl. Acad. Sci. U.S.A {\bf 95}, 98 (1998).

\bibitem{Kitaev2006}
A.~Kitaev,
\newblock Ann. Phys. {\bf 321}, 2 (2006).

\bibitem{Nayak2008}
C.~Nayak, S.~H.~Simon, A.~Stern, M.~Freedman, and S.~{Das Sarma},
\newblock Rev. Mod. Phys. {\bf 80}, 1083 (2008).



\bibitem{Motrunich2001}
O~ Motrunich, K.~Damle, and D.~A.~Huse, \newblock Phys. Rev. B
{\bf 63}, 224204 (2001).

\bibitem{Brouwer2011} 
P.~W.~Brouwer, M.~Duckheim, A.~Romito, and F.~von Oppen, \newblock
Phys. Rev. Lett. {\bf 107}, 196804 (2011).


 \bibitem{SGangadharaiahPRL2011}
S.~Gangadharaiah, B.~Braunecker, P.~Simon, and D.~Loss, Phys. Rev. Lett. \textbf{107}, 036801
  (2011).
\bibitem{LFidkowskiPRB2011c}
L.~Fidkowski, R.~M. Lutchyn, C.~Nayak, and M.~P.~A. Fisher,  Phys. Rev. B \textbf{84}, 195436 (2011).
  
  \bibitem{EMStoudenmirePRL2011}
E.~M. Stoudenmire, J.~Alicea, O.~A. Starykh, and M.~P.~A. Fisher,
   Phys. Rev. B \textbf{84}, 014503 (2011).

 \bibitem{ESelaPRB2011}
E.~Sela, A.~Altland, and A.~Rosch, Phys. Rev. B \textbf{84}, 085114 (2011).

\bibitem{GEVolovik2003}
G.~E. Volovik, Oxford University
  Press, 2003.
\bibitem{SDSarmaPRB2006}
S.~Das Sarma, C.~Nayak, and S.~Tewari,  Phys. Rev. B \textbf{73}, 220502
  (2006).



\bibitem[{\citenamefont{Fu and Kane}(2008)}]{Fu2008}
\bibinfo{author}{\bibfnamefont{L.}~\bibnamefont{Fu}} \bibnamefont{and}
  \bibinfo{author}{\bibfnamefont{C.~L.} \bibnamefont{Kane}},
  \bibinfo{journal}{Phys. Rev. Lett.} \textbf{\bibinfo{volume}{100}},
  \bibinfo{pages}{96407} (\bibinfo{year}{2008}).

\bibitem[{\citenamefont{Lutchyn et~al.}(2010)\citenamefont{Lutchyn, Sau, and
  {Das Sarma}}}]{Lutchyn2010}
\bibinfo{author}{\bibfnamefont{R.}~\bibnamefont{Lutchyn}},
  \bibinfo{author}{\bibfnamefont{J.}~\bibnamefont{Sau}}, \bibnamefont{and}
  \bibinfo{author}{\bibfnamefont{S.}~\bibnamefont{{Das Sarma}}},
  \bibinfo{journal}{Phys. Rev. Lett.} \textbf{\bibinfo{volume}{105}},
  \bibinfo{pages}{1} (\bibinfo{year}{2010}).

\bibitem[{\citenamefont{Oreg et~al.}(2010)\citenamefont{Oreg, Refael, and von
  Oppen}}]{Oreg2010}
\bibinfo{author}{\bibfnamefont{Y.}~\bibnamefont{Oreg}},
  \bibinfo{author}{\bibfnamefont{G.}~\bibnamefont{Refael}}, \bibnamefont{and}
  \bibinfo{author}{\bibfnamefont{F.}~\bibnamefont{von Oppen}},
  \bibinfo{journal}{Phys. Rev. Lett.} \textbf{\bibinfo{volume}{105}},
  \bibinfo{pages}{177002} (\bibinfo{year}{2010}).
 

\bibitem[{\citenamefont{Mourik et~al.}(2012)\citenamefont{Mourik, Zuo, Frolov,
  Plissard, Bakkers, and Kouwenhoven}}]{Mourik2012}
\bibinfo{author}{\bibfnamefont{V.}~\bibnamefont{Mourik}},
  \bibinfo{author}{\bibfnamefont{K.}~\bibnamefont{Zuo}},
  \bibinfo{author}{\bibfnamefont{S.~M.} \bibnamefont{Frolov}},
  \bibinfo{author}{\bibfnamefont{S.~R.} \bibnamefont{Plissard}},
  \bibinfo{author}{\bibfnamefont{E.~P. a.~M.} \bibnamefont{Bakkers}},
  \bibnamefont{and} \bibinfo{author}{\bibfnamefont{L.~P.}
  \bibnamefont{Kouwenhoven}}, \bibinfo{journal}{Science (New York, N.Y.)}
  \textbf{\bibinfo{volume}{336}}, \bibinfo{pages}{1003} (\bibinfo{year}{2012}).


\bibitem[{\citenamefont{Das et~al.}(2012)\citenamefont{Das, Ronen, Most, Oreg,
  Heiblum, and Shtrikman}}]{Das2012}
\bibinfo{author}{\bibfnamefont{A.}~\bibnamefont{Das}},
  \bibinfo{author}{\bibfnamefont{Y.}~\bibnamefont{Ronen}},
  \bibinfo{author}{\bibfnamefont{Y.}~\bibnamefont{Most}},
  \bibinfo{author}{\bibfnamefont{Y.}~\bibnamefont{Oreg}},
  \bibinfo{author}{\bibfnamefont{M.}~\bibnamefont{Heiblum}}, \bibnamefont{and}
  \bibinfo{author}{\bibfnamefont{H.}~\bibnamefont{Shtrikman}},
  \bibinfo{journal}{Nature Physics} \textbf{\bibinfo{volume}{8}},
  \bibinfo{pages}{887} (\bibinfo{year}{2012}).

\bibitem{Tewari2011}
S.~Tewari and  J.~D.~Sau \newblock Phys. Rev. Lett. {\bf 109}, 150408 (2012). 
  
  
  \bibitem{GKellsPRB2012a}
G.~Kells, D.~Meidan, and P.~W. Brouwer,  Phys. Rev. B \textbf{85}, 060507 (2012).

\bibitem[{\citenamefont{Fidkowski and Kitaev}(2010)}]{Fidkowski2010}
\bibinfo{author}{\bibfnamefont{L.}~\bibnamefont{Fidkowski}} \bibnamefont{and}
  \bibinfo{author}{\bibfnamefont{A.}~\bibnamefont{Kitaev}},
  \bibinfo{journal}{Phys. Rev. B} \textbf{\bibinfo{volume}{81}},
  \bibinfo{pages}{134509} (\bibinfo{year}{2010}).

\bibitem[{\citenamefont{Turner et~al.}(2011)\citenamefont{Turner, Pollmann, and
  Berg}}]{Turner2011}
\bibinfo{author}{\bibfnamefont{A.}~\bibnamefont{Turner}},
  \bibinfo{author}{\bibfnamefont{F.}~\bibnamefont{Pollmann}}, \bibnamefont{and}
  \bibinfo{author}{\bibfnamefont{E.}~\bibnamefont{Berg}},
  \bibinfo{journal}{Phys. Rev. B} \textbf{\bibinfo{volume}{83}},
  \bibinfo{pages}{075102} (\bibinfo{year}{2011}).

\bibitem[{\citenamefont{Fidkowski and Kitaev}(2011)}]{Fidkowski2011a}
\bibinfo{author}{\bibfnamefont{L.}~\bibnamefont{Fidkowski}} \bibnamefont{and}
  \bibinfo{author}{\bibfnamefont{A.}~\bibnamefont{Kitaev}},
  \bibinfo{journal}{Phys. Rev. B} \textbf{\bibinfo{volume}{83}},
  \bibinfo{pages}{1} (\bibinfo{year}{2011}).

\bibitem[{\citenamefont{Gurarie}(2011)}]{Gurarie2011}
\bibinfo{author}{\bibfnamefont{V.}~\bibnamefont{Gurarie}},
  \bibinfo{journal}{Phys. Rev. B} \textbf{\bibinfo{volume}{83}},
  \bibinfo{pages}{085426} (\bibinfo{year}{2011}).

\bibitem[{\citenamefont{Manmana et~al.}(2012)\citenamefont{Manmana, Essin,
  Noack, and Gurarie}}]{Manmana2012}
\bibinfo{author}{\bibfnamefont{S.}~\bibnamefont{Manmana}},
  \bibinfo{author}{\bibfnamefont{A.}~\bibnamefont{Essin}},
  \bibinfo{author}{\bibfnamefont{R.}~\bibnamefont{Noack}}, \bibnamefont{and}
  \bibinfo{author}{\bibfnamefont{V.}~\bibnamefont{Gurarie}},
  \bibinfo{journal}{Phys. Rev. B} \textbf{\bibinfo{volume}{86}},
  \bibinfo{pages}{205119} (\bibinfo{year}{2012}).

\bibitem{note1} {To be precise, 
Ref.\ \onlinecite{Fidkowski2010} proves the equivalent statement that the classes with $n=0$ and $n=8$ are 
adiabatically connected. An example of a continuous connection between $n=4$ and $n=-4$ is given in 
this article.}

\bibitem{Law2009}
K.~T.~Law, P.~A.~Lee, and T.~K.~Ng, Phys. Rev. Lett. {\bf 103}, 237001 (2009).

\bibitem{Flensberg2010}
K.~Flensberg, Phys. Rev. B {\bf 82}, 180516(R) (2010). 

\bibitem{note2} {Higher
values of the Andreev conductance are possible in principle, but these are unstable to perturbations
that preserve the effective time reversal symmetry.}

\bibitem[{\citenamefont{Fulga et~al.}(2011)\citenamefont{Fulga, Hassler,
  Akhmerov, and Beenakker}}]{Fulga2011}
\bibinfo{author}{\bibfnamefont{I.~C.} \bibnamefont{Fulga}},
  \bibinfo{author}{\bibfnamefont{F.}~\bibnamefont{Hassler}},
  \bibinfo{author}{\bibfnamefont{A.~R.} \bibnamefont{Akhmerov}},
  \bibnamefont{and} \bibinfo{author}{\bibfnamefont{C.~W.~J.}
  \bibnamefont{Beenakker}} (\bibinfo{year}{2011}), \eprint{arXiv:1101.1749v2}.



\bibitem[{\citenamefont{Kitaev}(2001)}]{Kitaev2001}
\bibinfo{author}{\bibfnamefont{A.~Y.} \bibnamefont{Kitaev}},
  \bibinfo{journal}{Phys. Usp.} \textbf{\bibinfo{volume}{44}},
  \bibinfo{pages}{131} (\bibinfo{year}{2001}).


\bibitem[{\citenamefont{Altland and Zirnbauer}(1997)}]{Altland1997}
\bibinfo{author}{\bibfnamefont{A.}~\bibnamefont{Altland}} \bibnamefont{and}
  \bibinfo{author}{\bibfnamefont{M.~R.} \bibnamefont{Zirnbauer}},
  \bibinfo{journal}{Phys. Rev. B} \textbf{\bibinfo{volume}{55}},
  \bibinfo{pages}{1142} (\bibinfo{year}{1997}).

\bibitem[{\citenamefont{Rieder et~al.}(2013)\citenamefont{Rieder, Brouwer, and
  I. Adagideli}}]{Rieder2013}
\bibinfo{author}{\bibfnamefont{M.-T.} \bibnamefont{Rieder}},
  \bibinfo{author}{\bibfnamefont{P.~W.} \bibnamefont{Brouwer}},
  \bibnamefont{and}
  \bibinfo{author}{\bibfnamefont{Ä.}~\bibnamefont{Adagideli}},
  \bibinfo{journal}{Phys. Rev. B} \textbf{\bibinfo{volume}{88}},
  \bibinfo{pages}{060509} (\bibinfo{year}{2013}).


\bibitem[{Note1()}]{Note1} \bibinfo{note}{This is in contrast to recent theoretical works that
   considered a wire coupled to a superconducting island, where non-local
   charging effects couple Majorana modes at opposite ends of the wire giving rise to a topological Kondo effect.~\cite
   {Altland2013,Beri2012}.}
 \bibitem[{\citenamefont{B\'{e}ri and Cooper}(2012)}]{Beri2012}
 \bibinfo{author}{\bibfnamefont{B.}~\bibnamefont{B\'{e}ri}} \bibnamefont{and}
   \bibinfo{author}{\bibfnamefont{N.~R.} \bibnamefont{Cooper}},
   \bibinfo{journal}{Phys. Rev. Lett.} \textbf{\bibinfo{volume}{109}},
   \bibinfo{pages}{156803} (\bibinfo{year}{2012}).



 \bibitem[{\citenamefont{Altland and Egger}(2013)}]{Altland2013}
 \bibinfo{author}{\bibfnamefont{A.}~\bibnamefont{Altland}} \bibnamefont{and}
   \bibinfo{author}{\bibfnamefont{R.}~\bibnamefont{Egger}},
   \bibinfo{journal}{Phys. Rev. Lett.} \textbf{\bibinfo{volume}{110}},
   \bibinfo{pages}{196401} (\bibinfo{year}{2013}).



\bibitem{foot3} {Note 
that it is the effective time reversal symmetry in the original formulation of the problem that 
protects this symmetric point, and that no fine tuning is needed to arrive at the Hamiltonian
(\ref{eq:HamSymmAnd}).}

\bibitem[{\citenamefont{Ng and Lee}(1988)}]{Ng1988}
\bibinfo{author}{\bibfnamefont{T.~K.} \bibnamefont{Ng}} \bibnamefont{and}
  \bibinfo{author}{\bibfnamefont{P.~A.} \bibnamefont{Lee}},
  \bibinfo{journal}{Phys. Rev. Lett.} \textbf{\bibinfo{volume}{61}},
  \bibinfo{pages}{1768} (\bibinfo{year}{1988}).














  

   






\end{thebibliography}

\section{Supplementary Material  }
\subsection{Scattering matrix of  a Majorana chain with topological index $|n|>4$}

Upon including interactions, the free-fermion cases with $|n| = 5$, $6$, $7$, or $8$ 
can be continuously connected to free-fermion classes with $|n| = 3$, $2$, $1$, and $0$,
respectively. For the cases $n = \pm 5$, $\pm 6$, or $\pm 7$ this requires the
addition of extra channels (at least, locally, near the end of the Majorana chain), similarly to 
the transition from $n = 4$ to $n=-4$ discussed in the main text. For presentation purposes we focus 
on the case $n=7$; the other seven cases have similar phenomenology.

\begin{figure}
\includegraphics[width=0.45\textwidth]{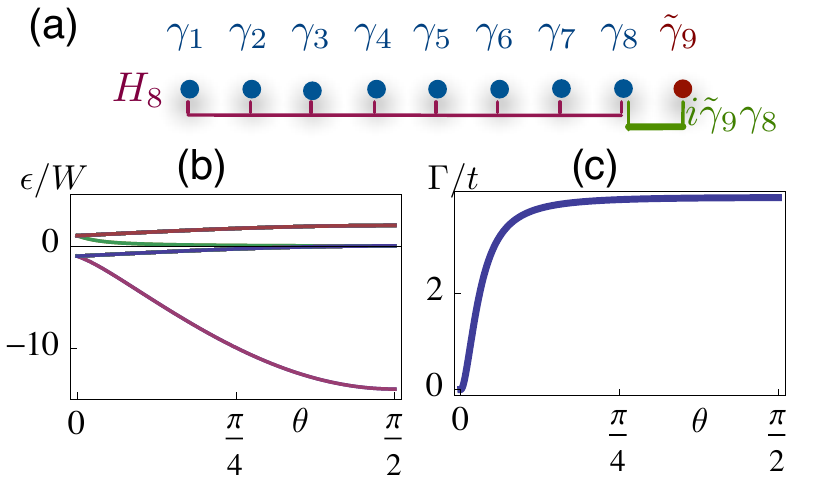}
\caption{(a) Schematic picture a nine channel setup that allows for an interpolation between the 
topologically equivalent effective free-fermion phases with $n=7$ and $n=-1$. 
(b) Energy spectrum of the low energy excitations as a function of the interpolation parameter
$\theta$, showing that the system remains gapped throughout the transition. (c) Modulus square
of single-particle tunneling matrix elements between the two degenerate ground states, as a function 
of  $\theta$.\label{fig1}}
\end{figure}

Consider the nine-channel setup depicted in Fig.~\ref{fig1}. The channels $\alpha=1,2,\ldots,8$ have
positive $\Delta$, whereas channel $\alpha=9$ has negative $\Delta$. Without interactions, the
topological index $n = N - \tilde N = 7$. To the non-interacting 
Hamiltonian $H$ of Eq. 2 in the main text 
we add the perturbation:
\begin{equation}
  H'(\theta) =  W ( i \cos \theta \tilde{\gamma}_9 \gamma_8 + H_8 \sin \theta)
  \label{eq:H7to-1}
\end{equation}
where the interaction $H_8$ is given by Eq. 8 in the main text. 
For $0 < \theta \le \pi/2$ the ground
state is twofold degenerate. For $\theta = 0$ the two Majorana end states $\gamma_8 $ and 
$\tilde{\gamma}_9 $ gap out,  and the system is effectively in the free-fermion $n=7$ configuration.
At this special point one expects $\mbox{tr}\, r_{\rm he} = 7$.
Conversely, when $\theta = \pi/2$, the interaction term $H_8$ renders the eight first channels 
trivial, leaving a single Majorana state $\tilde \gamma_9$, corresponding to the free-fermion $n=-1$ 
configuration with $\mbox{tr}\, r_{\rm he} = -1$. An explicit diagonalization of Eq.~\eqref{eq:H7to-1} 
shows that the gap remains open throughout the interpolation between these two extremes, see Fig.\ 
\ref{fig1}, indicating that the two are indeed in the same topological phase.

We now consider how  $r_{\rm he}$ goes between its two extreme values 
if $\theta$ is taken from $0$ to $\pi/2$. For $\theta$ small but positive the 
interaction $H_8$ lifts the ground state degeneracy of the free-fermion $n=7$ configuration, up to a 
twofold degeneracy involving two states with opposite fermion parity, which are simultaneously located at 
both ends of the superconducting wire.
An explicit calculation shows that single-particle tunneling events do not couple the two ground states 
in the limit $\theta \rightarrow 0^+$, see Fig.~\ref{fig1}(c), indicating that the end states are not 
of single particle nature in this limit. Instead, in the limit $\theta \rightarrow 0^+$ transitions between 
the two ground states require the exchange of multiple fermions with the leads. In contrast, for generic
$0 < \theta \le \pi/2$ the end spectrum is essentially of single particle nature, 
and the exchange of a single fermion with the leads is sufficient for a transition between the two degenerate
ground states. The degeneracy, fermion parity, and transition matrix elements between the states at 
generic $0 < \theta \le \pi/2$ evolve continuously into the free-fermion state at $\theta=\pi/2$, which
has a single Majorana state of ``$\tilde \gamma$'' type at the end of the Majorana chain.
In the weak coupling limit $|t_{\rm T}| \ll |t|$ such single particle tunneling events dominate over 
the higher order processes which derive from the limit $\theta \rightarrow 0$, so that for weak
coupling the system behaves as a $n=-1$ free fermion configuration along the entire path $0 
< \theta \le \pi/2$, with the exception of the special point $\theta = 0$. We conclude that in a generic
setting the free-fermion $n=7$ configuration is unstable to interactions, and that in the 
weak-coupling limit the interacting system  flows to an effectively non-interacting 
configuration with $\mbox{tr}\, r_{\rm he} = -1$.

A similar analysis can be applied to the remaining cases $|n|=5$, $|n|=6$, and $|n|=8$,
and shows that upon inclusion of interactions these configurations are dominated by effectively
free-fermion configurations with $|n|=3$, $|n|=2$, and $|n|=0$, respectively.


\end{document}